\documentstyle[12pt]{article}

\newcommand{\bce}{\begin{center}}
\newcommand{\ece}{\end{center}}
\newcommand{\be}{\begin{equation}}
\newcommand{\ee}{\end{equation}}
\newcommand{\bea}{\begin{eqnarray}}
\newcommand{\eea}{\end{eqnarray}}

\newcommand{\ba}{\begin{array}}
\newcommand{\ea}{\end{array}}

\newcommand{\bsigma}{\mbox{\boldmath $\sigma$}}

\def\lsim{\mathrel{\rlap{\lower4pt\hbox{\hskip1pt$\sim$}}
    \raise1pt\hbox{$<$}}}         
\def\gsim{\mathrel{\rlap{\lower4pt\hbox{\hskip1pt$\sim$}}
    \raise1pt\hbox{$>$}}}         

\def\Pom{{\bf I\!P}}

\def\lsim{\mathrel{\rlap{\lower4pt\hbox{\hskip1pt$\sim$}}
    \raise1pt\hbox{$<$}}}         

\def\gsim{\mathrel{\rlap{\lower4pt\hbox{\hskip1pt$\sim$}}
    \raise1pt\hbox{$>$}}}         

\begin{document}
\begin{center}
\phantom{.}{\large\bf FZJ-IKP(Th)-08/99 }
\vspace{2.0cm}\\
{\Large \bf Diffraction driven steep rise of spin structure function
$g_{LT}=g_{1}+g_{2}$ at small x and DIS sum rules}
\vspace{2.0cm}\\
{\large  I.P.Ivanov$^{1,2)}$, N.N. Nikolaev $^{1,3)}$,  A.V.
Pronyaev$^{4)}$ 
and W.Sch\"afer$^{1)}$\vspace{0.5cm}\\}
{\sl
$^{1}$IKP(Theorie), KFA J{\"u}lich, D-52428 J{\"u}lich, Germany
\medskip\\
$^{2)}$Novosibirsk University, Novosibirsk, Russia\medskip\\
$^{3}$L. D. Landau Institute for Theoretical Physics, GSP-1,
117940, \\
ul. Kosygina 2, Moscow 117334, Russia\\
$^{4}$Virginia Polytechnic Institute and State University, 
Blacksburg, VA 24061-0435, USA}
\vspace{1.0cm}


{\bf Abstract}
\end{center}

We derive a unitarity relationship between the spin structure 
function $g_{LT}(x,Q^{2})=g_{1}(x,Q^{2})+g_{2}(x,Q^{2})$, the LT 
interference diffractive structure function and the spin-flip
coupling of the pomeron to nucleons. Our diffractive mechanism
gives rise to a dramatic small-$x$ rise $g_{LT}(x,Q^{2})
\sim g^{2}(x,\overline{Q}^{2}) 
\sim \left({1\over x}\right)^{2(1+\delta_{g})}$, 
where $\delta_{g}$ is an exponent of small-$x$ rise of the 
unpolarized gluon density in the proton $g(x,\overline{Q}^{2})$ at a
moderate hard scale $\overline{Q}^{2}$ for light flavour contribution 
and large hard scale $\overline{Q}^{2} \sim m_{f}^{2}$ for heavy flavour 
contribution. It invalidates the Burkhardt-Cottingham sum rule. The
found small-$x$ rise of diffraction driven 
$g_{LT}(x,Q^{2})$ is steeper than given by the
Wandzura-Wilczek relation under conventional assumptions on small-$x$
behaviour of $g_{1}(x,Q^{2})$.
\vspace{1.0cm}\\

\section{Introduction and motivation}

The combination $g_{LT}(x,Q^{2})=g_{1}(x,Q^{2})+g_{2}(x,Q^{2})$ of familiar 
spin structure functions $g_{1}$ and $g_{2}$ of deep inelastic scattering
(DIS) is related to the absorptive 
part of amplitude $A_{\mu\rho,\nu\lambda}({\bf \Delta}=0)$ of 
forward $(T)$ transverse to $(L)$ longitudinal photon scattering 
accompanied by the target nucleon spin-flip,
\begin{equation}
\sigma_{LT} = {1\over (Q^{2}+W^{2})}{\rm Im}  A_{-1-{1\over2},
L+{1\over 2}}({\bf\Delta}=0) = {4 \pi ^{2} \alpha_{em} \over Q^{2}}
\cdot {4m_{p} \over \sqrt{Q^{2}}}\cdot x^{2}g_{LT}(x,Q^{2})\, ,
\label{eq:1}
\end{equation}
where $\bf{\Delta}$ is the  momentum transfer and $\mu,\nu=\pm 1,L$ and
$\rho,\lambda= \pm {1\over 2}$ are helicities of particles in 
$\gamma^{*}_{\nu} p_{\lambda} \to \gamma^{*}_{\mu}p'_{\rho}$ scattering,
$Q^{2},W^{2}$ and $x=Q^{2}/(Q^{2}+W^{2})$ are standard DIS variable.
The motto of high energy QCD -- the quark helicity conservation, 
the common wisdom that high energy scattering is spin-independent, some 
model considerations including 
\cite{BC} the vanishing one-pomeron exchange contribution to 
$A_{-1-{1\over2},L+{1\over 2}}({\bf\Delta}=0)$ , all suggest that the 
corresponding spin asymmetry $A_{2}=\sigma_{LT}/\sigma_{T}$  vanishes 
in small-$x$ limit of DIS.

In this communication we demonstrate that this is not the case. We find
about $x$-independent  spin asymmetry $A_{2}$ and scaling and steeply rising  
$g_{LT}(x,Q^{2})$ at small $x$,
\begin{equation}
g_{LT}(x,Q^{2})\sim {G^{2}(x,{\overline Q}^{2})\over x^{2}}
  , 
\label{eq:2}
\end{equation}
where $G(x,Q^{2})=xg(x,Q^{2})\sim \left({1\over x}\right)^{\delta_{g}}$ 
is the conventional unpolarized gluon structure function of the target 
nucleon and  ${\overline Q}^{2}$ is flavour dependent scale to be specified
below. 

The case of the helicity amplitude $A_{-1-{1\over2},L+{1\over 2}}
({\bf\Delta})$ is quite tricky. On the one hand,  QCD motivated 
considerations strongly suggest a nonvanishing pomeron 
spin-flip in diffractive  nucleon-nucleon scattering \cite{Zakharov}. On 
the other hand, recent studies have shown that the $s$-channel helicity 
nonconserving (SCHNC) LT interference cross section $\sigma_{LT}^{D}$ of
diffractive DIS \cite{NPLT} and related SHCNC spin-flip amplitudes of 
diffractive vector meson production do not vanish \cite{KNZ98,IN99} at 
small $x$. As Zakharov emphasized \cite{Zakharov} such spin-flip does not
conflict the quark helicity conservation because in scattering of 
composite objects helicity of composite states is not equal to the sum 
of helicities of quarks, which arguably holds way beyond the perturbative 
QCD (pQCD) domain.  The recent work on SCHNC vector meson production 
illustrates this point  nicely \cite{NPLT,KNZ98,IN99}. 

Consequently, pomeron exchange well contributes to this helicity amplitude
but the Procrustean bed of Regge factorization enforces the forward zero, 
$A_{-1-{1\over2},L+{1\over 2}}({\bf\Delta}) \propto {\bf \Delta}^{2}$, and
vanishing $\sigma_{LT}$ in one-pomeron exchange approximation. 
The principal point behind our result (\ref{eq:2}) is Gribov's observation
\cite{Gribov} that such kinematical zeros can be lifted by two-pomeron 
exchange (two-pomeron cut) which can contribute to helicity amplitudes 
vanishing in 
one-pomeron exchange approximation. A good example is a recent derivation 
\cite{NNSchaefer} of a rising tensor structure function $b_{2}(x,Q^{2})$ 
for DIS off spin-1 deuterons. In defiance of common wisdom, it gives rise 
to dependence of total cross section on the deuteron tensor polarization 
which persists at small $x$. Such a rise of  $b_{2}(x,Q^{2})$ 
invalidates the Close-Kumano sum rule \cite{CloseKumano}. Incidentally, 
it derives for the most part from diffractive mechanism which we pursue 
in this paper. Another example due to Karnakov \cite{Karnakov} is a
difference of $\gamma\gamma$ total cross sections for parallel and 
perpendicular linear polarizations of colliding photons - the quantity 
which vanishes in one-pomeron exchange approximation. The keyword 
behind these new effects is unitarity \cite{AGK}, two-pomeron cut is 
simply a first approximation to imposition of unitarity constraints. 
 
\section{Regge theory expectations and sum rules}
   
We recall that our expectations for small-$x$ behaviour of different 
structure functions, $\sim \left({1\over x}\right)^{\delta}$, have been 
habitually driven by the Regge picture of soft interactions, in which the 
exponent (intercept) $\delta=\alpha-1$ is controlled by quantum numbers 
of the relevant $t$-channel exchange (a good summary is found in 
textbook \cite{Ioffe}). For instance, the Regge theory suggests 
$\alpha_{\Pom} \sim 1$ for helicity-diagonal pomeron (vacuum) exchange 
dominated $F_{2}(x,Q^{2})$ and $F_{L}(x,Q^{2})$  and $\alpha_{R} \sim 
{1\over 2}$ for the secondary reggeon $(\rho,A_{2})$-exchange quantities 
like $F_{2p}(x,Q^{2})-F_{2n}(x,Q^{2})$ and $\omega$-exchange $F_{3}(x,Q^{2})$.
The dominant $A_{1}$ and $f_{1}$ reggeon exchange in the axial vector 
channel suggests $\alpha_{1} \sim 0$ for $xg_{1}(x,Q^{2})$. These Regge 
theory intercepts are not stable against QCD evolution, but extensive 
studies of small-$x$ asymptotics of generalized two-gluon and 
quark-antiquark ladder diagrams have revealed only marginal modifications 
of the above hierarchy of intercepts (for the BFKL pomeron exchange see 
\cite{BFKL}, for reggeon exchange and/or non-singlet structure function 
see \cite{ErmolaevVal}, for different spin structure functions see: 
$g_{1}(x,Q^{2})$ in \cite{ErmolaevG1}, $g_{2}(x,Q^{2})$ in \cite{ErmolaevG2}, 
$F_{3\gamma}(x,Q^{2})$ in \cite{ErmolaevF3}). The corollary of these 
studies is that $g_{1}(x,Q^{2})$ and $g_{2}(x,Q^{2})$ of two-parton 
ladder approximation have the $x$-dependence typical of the reggeon 
exchange and  their contributions to spin asymmetries $A_{1}$ and $A_{2}$ 
do indeed vanish in the small-$x$ limit. We recall that works 
\cite{ErmolaevG1,ErmolaevG2,ErmolaevF3} focused on exactly forward, 
${\bf \Delta}=0$, Compton scattering amplitudes. 

Burkhardt and Cottingham \cite{BC} argued that because neither pomeron nor 
high lying reggeon exchanges contribute to $A_{-1-{1\over2},L+{1\over 2}}
({\bf\Delta}=0)$, then unsubtracted (superconvergent) dispersion relation 
holds for this Compton scattering amplitudes. Precisely superconvergence 
has been the principal assumption behind the much discussed BC sum rule 
\cite{BC}
\begin{equation}
\int dx g_{2}(x,Q^{2})\propto \int_{Q^{2}/2}^{\infty} d\nu
{\rm Im} A_{2}(Q^{2},\nu,{\bf \Delta}=0) =0\, ,
\label{eq:3}
\end{equation}
for thorough reviews see \cite{Ioffe,Jaffe,Anselmino,Reya}. The tricky
point is that the BC amplitude $A_{2}(Q^{2},\nu,{\bf \Delta})$ (which 
differs from our $A_{-1-{1\over2},L+{1\over 2}}({\bf \Delta})$ only 
insignificantly) receives a contribution from pomeron exchange, and the 
integral 
$$
\int_{Q^{2}/2}^{\infty} d\nu{\rm Im} A_{2}(Q^{2},\nu,{\bf \Delta})
$$
would diverge at any finite ${\bf \Delta\neq 0}$, which makes the BC sum 
rule quite a singular one. As we emphasized above, $A_{-1-{1\over2},
L+{1\over 2}}({\bf\Delta}) \propto {\bf \Delta}^{2}$ and vanishes at 
${\bf \Delta =0}$ 
only because of rigours of  Regge factorization, Gribov's two-pomeron
exchange breaks Regge factorization and gives $A_{-1-{1\over2},L+{1\over 2}}
({\bf\Delta}=0) \neq 0$, see also \cite{Heimann}. The specter of resulting
dramatic small-$x$ rise of $g_{2}(x,Q^{2})$ and of divergence of the BC
integral permeates the modern literature on spin structure functions (see 
textbook \cite{Ioffe} and recent reviews \cite{Jaffe,Anselmino,Reya}).
The aforementioned breaking of the Close-Kumano some rule is of the same 
origin; if cast in the Regge language, the scaling and rising tensor 
structure function found in \cite{NNSchaefer} falls into the pomeron-cut 
category. Numerical estimates show that tensor asymmetry is quite large,
the related evaluations of $g_{LT}(x,Q^{2})$ are as yet lacking. 

\section{Diffractive DIS and unitarity driven $g_{LT}^{U}(x,Q^{2})$}

In this communication we fill this gap and report the first ever 
evaluation of unitarity or diffractive driven contribution to  
$g_{LT}(x,Q^{2})$ in terms
of the two other experimentally accessible spin observables: the SCHNC 
LT interference diffractive DIS structure function \cite{NPLT} and the 
pomeron spin-flip amplitude in nucleon-nucleon and pion-nucleon 
scattering \cite{Zakharov}. By unitarity relation, the opening of 
diffractive DIS channel $\gamma^{*}p\to p'X$ affects 
the elastic scattering amplitude 
(\cite{AGK} and references therein). The best known unitarity effect is 
Gribov's absorption or shadowing correction \cite{GribovShad} to one-pomeron
exchange. Besides simple shadowing, for spinning particles unitarity
corrections  can give rise to new spin amplitudes absent in one-pomeron 
exchange, which was precisely the case with tensor structure function 
for DIS off spin-1 deuteron \cite{NNSchaefer}. 

In the related evaluation of unitarity driven $\sigma_{LT}^{(U)}$ 
we start with the eikonal unitarity diagram in fig.~1. Here the
eikonal refers to the `elastic' $pX$ intermediate state, the effect 
of so-called `inelastic' intermediate states $p^{*}X$  when
the proton excites into resonances or low-mass continuum states 
will be commented on below. Hereafter all unitarity corrections 
will be supplied by a superscript $(U)$. As an 
input we need amplitudes $A_{\mu\rho,\nu\lambda}^D$ of diffractive 
DIS $\gamma^{*}_{\nu} p_{\lambda} \to X_{\mu}p'_{\rho}$, where $\mu$
stands for spin states of the diffractive state $X$. Applying the 
optical theorem to this unitarity contribution to forward scattering
amplitude, we find \cite{AGK}
\begin{equation}
\sigma_{LT}^{(U)} = {\rm Re } {1 \over 16\pi^{2}(W^{2}+Q^{2})^{2}} \int 
d^{2}{\bf \Delta} dM^{2}
\sum_{\mu,\rho} A^D_{-1-{1\over 2},\mu\rho}
A^D_{\mu\rho,L+{1\over 2}}\, ,
\label{eq:4}
\end{equation}
where $M$ is the invariant mass of the intermediate state. In order for
this unitarity diagram to contribute to $\sigma_{LT}$, the {\sl r.h.s.}
of (\ref{eq:4}) must have a structure which in the convenient polarization 
vector-spinor representation has the form  
\begin{equation}
\sigma_{LT} \propto \langle f|({\bsigma}[{\bf n e}^{\dagger}(-)])
|in\rangle \, ,
\label{eq:5}
\end{equation}
where $\bsigma$ is the nucleon spin operator, ${\bf n}$ is the 
unit vector along the $\gamma^{*}p$ collision axis, and 
$$
{\bf e}(\nu)= -{1 \over \sqrt{2}}(\nu,i)
$$ 
is the photon polarization vector for helicity $\nu$. 

In the polarization-vector representation the factorized one-pomeron 
amplitude for diffractive DIS reads \cite{KNZ98,IN99}
\begin{eqnarray}
A^{D}=(i+{\pi \over 2}\delta_{\Pom})
\left\{T_{0L}V^{\dagger}_{0}e_{L}+T_{\pm\pm} ({\bf V}^{\dagger}{\bf e})+ 
T_{\pm 0}V^{\dagger}_{0}({\bf \Delta}{\bf e})
+ T_{\pm L}({\bf \Delta}{\bf V}^{\dagger})e_{L} + ...\right\}\nonumber\\
\times 
\{1+i{r_{5}\over m_{p}}
{(\bsigma}[\bf{n\Delta}])\}\, ,
\label{eq:6}
\end{eqnarray}
where $T_{\mu\nu}$ is the imaginary part of diffractive amplitude for an 
unpolarized target, ${\bf V}$ stands for the transverse polarization vector
of diffractive state $X$  and $r_{5}=r_{5}(0)\exp(-B_{5}{\bf \Delta}^{2})$ 
is the ratio of the spin-flip to non-flip pomeron-nucleon couplings. The 
signature factor 
\begin{equation}
\eta = i+{\pi\over 2}\delta_{\Pom}\, .
\label{eq:7}
\end{equation}
is defined through the $Q^{2}$- and $x$-dependent effective intercept  
\begin{equation}
\delta_{\Pom} = {d \log Im T(x,Q^{2}) \over d\log{1\over x}}\, ,
\label{eq:8}
\end{equation} 
which is the same as $\delta_{g}$ taken at a relevant average hard scale.
The sight difference of this scale and of $\delta_{\Pom}$ thereof 
for different helicity amplitudes can be neglected for the purposes
of our discussion and to a good approximation $r_{5}$ can be considered 
a real
valued quantity.

The Regge factorization (\ref{eq:6}) is equally applicable to 
one-pomeron exchange elastic scattering, in which case it dictates
(hereafter we suppress the helicity $(-)$) 
\begin{equation}
A_{-1-{1\over 2},L+{1\over 2}}({\bf\Delta}) \propto 
T_{-L}({\bf e^{\dagger}}
{\bf
\Delta})({\bsigma}[\bf{n\Delta}])\, ,
\label{eq:9}
\end{equation} 
which, as we mentioned in the Introduction, vanishes in the forward case 
${\bf \Delta}=0$.

We need the $LT$ transition in either of the diffractive $\gamma^{*}
\leftrightarrow X$ vertices and spin-flip transition in either of the 
pomeron-nucleon vertices in unitarity diagram of fig.~1, the other
two vertices are spin non-flip ones. The both spin-flip transitions are
off-forward with finite momentum transfer ${\bf \Delta}$ to the 
intermediate state and the integrand of eq.~(\ref{eq:4}) will
be $\propto ({\bf e}^{\dagger}{\bf \Delta})(\bsigma [{\bf n  \Delta}])$. 
Summing over the phase space of the
intermediate state $X$ includes an integration over azimuthal 
angle $\phi$ of ${\bf \Delta}$ of the form
\begin{equation}
\int {d\phi \over 2\pi}\,
({\bf e}^{\dagger}{\bf \Delta})(\bsigma [{\bf n  \Delta}])
={1\over 2}\Delta^{2}(\bsigma [\bf{n e}^{\dagger}])\, ,
\label{eq:10} 
\end{equation}
which has precisely the desired spin structure (\ref{eq:5}).

Now we notice that the LT interference term in differential cross section 
of diffractive DIS on unpolarized nucleons  
$\gamma^{*}_{\nu} p_{\lambda} \to X_{\mu}p'_{\rho}$ equals
\begin{equation}
{d\sigma^D_{LT} \over dM^{2} d^{2}{\bf \Delta}} = 
{1 \over 16\pi^{2} (W^{2}+Q^{2})^{2}} 
\sum_{\mu,\rho\lambda} A^{D*}_{-1\lambda,\mu\rho}
A^D_{\mu\rho,L\lambda}\, 
\label{eq:11}
\end{equation}
and differs from the {\sl r.h.s.} of (\ref{eq:4}) only by complex
conjugation of one of diffractive amplitudes.
In principle $\sigma_{LT}^{D}$ can be measured experimentally. The 
scaling properties of $d\sigma_{LT}^{D}$ have been established in 
\cite{NPLT}. The conventionally defined LT interference diffractive 
structure function $F_{LT}^{(4)}$ is twist-3 \cite{NPLT}, for the 
purposes of the present discussion it is convenient to factor out the 
kinematical factor $\Delta/Q$ and define the {\sl scaling} and 
{\sl dimensionless} LT diffractive structure function $g^{D}_{LT}(x_{\Pom},
\beta,Q^{2})$ such that
\begin{equation}
(M^{2}+Q^{2}) {\sigma^D_{LT} \over dM^{2}d{\bf \Delta}^{2}}
={4 \pi^{2}\alpha_{em} \over Q^{2}} \cdot {({\bf \Delta e}) \over Q}
\cdot \left(1+|r_{5}|^{2}{{\bf \Delta}^{2} \over m_{p}^{2}}\right)
\cdot g^{D}_{LT}(x_{\Pom},\beta,Q^{2}) B_{LT} \exp(-B_{LT}{\bf \Delta}^{2})\,,
\label{eq:12}
\end{equation}
where $\beta=Q^{2}/(Q^{2}+M^{2})$ and $x_{\Pom}=x/\beta$ are diffractive
DIS variables. In what follows we shall neglect corrections $\propto 
|r_{5}|^{2}$, because nucleon spin-flip 
effects are numerically very small within 
the diffraction cone. Then, making use of (\ref{eq:11}),(\ref{eq:12}) and 
of the factorization property of one-pomeron amplitude (\ref{eq:6}), we 
obtain 
\begin{equation}
g_{LT}^{(U)}(x,Q^{2}) = {1\over x^{2}}\cdot
 r_{5}(0)\sin(\pi \alpha_{\Pom})\cdot
\int_{x}^{1} {d\beta \over \beta}\cdot {B_{LT} \over 4m_{p}^{2}
(B_{LT}+B_{5})^{2}} \cdot g^{D}_{LT}(x_{\Pom},\beta,Q^{2})\, .
\label{eq:13}
\end{equation}

\section{The model evaluations of $g_{LT}^{(U)}$}

Eq.~(\ref{eq:13}) is our central result and up to now we have been 
completely model independent. In principle, the $g_{LT}^{D}$ can be
measured experimentally, in the lack of such direct data in our 
numerical estimates of $\sigma_{LT}$ we resort to QCD model for 
diffractive DIS developed in \cite{NPLT}. We refer to this paper for 
details, here we only recall the salient results.

The driving term of diffractive DIS is excitation of $q\bar{q}$ Fock 
states of the photon (fig.~2). 
We notice that only  $q\bar{q}$ pairs with the sum of helicities 
zero contribute to $\sigma_{LT}$. Consider first the contribution from 
intermediate heavy flavour excitation, in which case the mass $m_{f}$ 
of a heavy quark provides the large pQCD hard scale
\cite{GNZcharm,BGNPZ98}
\begin{equation}
\overline{Q}^{2}\approx {m_{f}^{2} \over 1-\beta}\, .
\label{eq:14}
\end{equation} 
The lower blobs in the diagram of fig.~2a are related to skewed unintegrated 
gluon structure function of the proton which can be approximated by the 
conventional diagonal unintegrated gluon structure function taken at 
$x_{eff} = {1\over 2}x_{\Pom}$. To a $\log\overline{Q}^{2}$ accuracy, 
gross features of $g^{D}_{LT}(x_{\Pom},\beta,Q^{2})$ are 
described by \cite{NPLT} 
\begin{eqnarray}
g^{D}_{LT}\approx { e_f^2 \over 3B_{LT} m_f^2}\cdot
\beta^4 (1-\beta)(2-3\beta)\alpha^2_{S}(\overline{Q}^{2})
G^{2}({1\over 2}x_{\Pom},\overline{Q}^{2})\, ,
\label{eq:15}
\end{eqnarray}
where $e_{f}$ is the quark charge in units of the electron charge, 
$\alpha_{S}$ is the strong coupling, and we assumed $Q^{2}\gg 4m_{f}^{2}$.
Notice that the hard scale (\ref{eq:14}) rises as $\beta \rightarrow 1$.
The QCD scaling violations in the gluon structure function are strong and
at moderate values of $\overline{Q}^{2}$ the crude approximation is
\begin{equation}
G^{2}({1\over 2}x_{\Pom},\overline{Q}^{2}) \propto
\overline{Q}^{2\gamma}\left({1\over x_{\Pom}}\right)^{2\delta_{g}}
\propto {\beta^{2\delta_{g}} \over (1-\beta)^{\gamma}}
\left({1\over x}\right)^{2\delta_{g}}\, ,
\label{eq:16} 
\end{equation}
where $\gamma \sim 1$ and $2\delta_{g} \sim$ 0.4-0.5 for moderate $Q^{2}$
and $x\lsim 10^{3}$, see below. 
Both the scaling violations and, to a lesser extent the small-$x$ rise of 
gluon densities, enhance the contribution from $\beta \sim 1$. Then the  
contribution from intermediate open heavy flavor to 
the {\sl l.h.s.} of (\ref{eq:13}) can be evaluated as
\begin{equation}
g_{LT}^{(U)}(x,Q^{2}) \approx -{1\over 30 x^{2}}
{r_{5}(0)e_{f}^{2} \over (B_{LT}+B_{5})^{2}2m_{p}^{2}m_{f}^{2}}\cdot
\alpha_{S}^{2}(\overline{Q}^{2}) 
G^{2}({1\over 2}x,\overline{Q}^{2}\approx m_{f}^{2})\, .
\label{eq:17}
\end{equation}
The numerical factor in the {\sl r.h.s.} of (\ref{eq:17}) is only a 
crude estimate for $\gamma\approx 1$, it depends strongly on the pattern 
of scaling violations which for heavy flavours is under the control of pQCD.

The detailed
discussion of pQCD hard scale for light flavour contribution to $g^D_{LT}$ 
is found in \cite{NPLT}. We only emphasize that because the dominant 
contribution to $\sigma_{LT}$ comes from $\beta \sim 1$ where the hard 
scale (\ref{eq:14}) is enhanced by the factor $\propto 1/{1-\beta}$, 
even for light flavours $\sigma_{LT}$ receives a dominant contribution 
from hard to semihard gluons. The final result for $g_{LT}^{D}$ is similar 
to (\ref{eq:15}) with the substitution of $m_{f}^{2}$ by the semihard 
scale $\overline{Q}^{2} \approx $(0.5-1) GeV$^{2}$. Still it is not under 
the full control of pQCD because of substantial contribution of soft 
momenta in the quark loop integration. Also, because $g_{LT}^{D}$
changes the sign, the numerical results for the moment (\ref{eq:13}) 
depend on the pattern of scaling violations at the semihard scale 
$\overline{Q}^{2} \approx $(0.5-1) GeV$^{2}$ which can be affected by
soft dynamics. Furthermore, the overall result for unitarity driven $g_{LT}$
is proportional to $\langle {\bf \Delta}^{2}\rangle$, which is the soft 
quantity controlled by the size of the target proton and large 
transverse size of diffractive $q\bar{q}$ states \cite{NPZslope}.
All this sensitivity 
to soft input notwithstanding, our unitarity driven $g_{LT}^{(U)}$ 
has precisely 
the same QCD status as standard diffractive structure functions: it exists, 
it is a scaling phenomenon, its QCD evolution is reasonably well understood
\cite{NZ94}, but its numerical magnitude  is not calculable from first 
principles of pQCD, although QCD motivated models do correctly reproduce 
all features of diffractive DIS \cite{BGNPZ98,GNZ95}. To this end we 
recall that no one has ever requested pQCD to provide the input for 
standard DGLAP evolution of the proton structure function.

The small-$x$ dependence of $x^{2}g_{LT}$ is the same \cite{NPLT} as of
the unpolarized diffractive structure function \cite{GNZcharm,BGNPZ98} 
and, in principle, could be
borrowed from experiment. The effective exponent $\delta_{g}$ depends 
on $x_{g}$, at $x_{g}=10^{-3}$ for light flavour contribution we find
$2\delta_{g} \approx 0.4$, for the numerically smaller charm
contribution $2\delta_{g} \approx  0.6.$. 

\section{Numerical estimates: unitarity driven $g_{LT}^{(U)}$ vs. the
Wandzura-Wilczek relationship}

The nucleon spin-flip defines a brand new skewed gluon distribution 
\cite{Radyushkin,Ji}, without going into details we only state that 
anomalous dimensions which control the small-$x$ dependence of this 
skewed structure function are identical to those for unpolarized gluon 
distribution and the spin-flip parameter $r_{5}$ would depend on neither
$x$ nor $Q^{2}$. 
Zakharov's sound arguments \cite{Zakharov} in favor of non-vanishing $r_{5}$ 
do not require pQCD and the existence of our unitarity 
driven $g_{LT}(x,Q^{2})$ is beyond doubts. However, as a soft parameter 
$r_{5}$ is quite sensitive to models of soft wave function of the nucleon 
\cite{Zakharov}. Incidentally, it is of great interest
for the polarimetry of stored proton beams and the whole spin physics 
program at RHIC \cite{Buttimore}.
Different model estimates of
$r_{5}$ and the experimental situation are summarized in recent review 
\cite{Buttimore}. The experimental data on pion-nucleon scattering 
give $|r_{5}| = 0.2\pm 0.3$, the experimental data on proton-proton 
scattering leave a room for quite a strong spin-flip, $r_{5}=-0.6$
with about 100 per cent uncertainty. The theoretical models give 
$|r_{5}| \lsim $ 0.1-0.2, the sign of $r_{5}$ remains open. 

Even if 
$r_5$ were known, there will be corrections to our eikonal estimate
(\ref{eq:13}) from proton excitations $p^*$ (resonances and continuum) 
in the intermediate state (fig.~1b). Although SCHC diffraction
excitation amplitudes are smaller than elastic ones 
(for suppression of diffraction excitation by the node effect see \cite{NNN}), 
the inelastic $(p,p^{*})$ spin-flip 
transitions can be enhanced compared to elastic $(p,p')$ ones 
\cite{IN99}. Consequently, one can not exclude that the contribution
of $p^{*}$ excitations would enhance the effective $r_{5}$ by a large
factor. Here for the sake of definiteness we 
evaluate $x^{2}g_{LT}^{D}(x,Q^{2})$ assuming the conservative value
$r_{5}=-0.1$. We take $B_{LT}=10$ GeV$^{2}$ for light flavours and
$B_{LT}=5$ GeV$^{-2}$ for charm  as evaluated in \cite{NPLT}, the slope
$B_{5}$ remains unknown and we put $B_{5}=0$. 
This conservative estimate is shown in fig.~3, at the moment we can 
not exclude even one order in magnitude larger effect. At $Q^{2} \lsim
4m_{c}^{2}$ the charm contribution to diffraction is small, the difference
between small-$Q^{2}$ and large-$Q^{2}$ curves in fig.~3 illustrates 
the significance of charm contribution to $g_{LT}$. A crude 
parameterization of our numerical results for $x\lsim 10^{-3}$ and
$Q^{2}=5$ GeV$^{2}$ is
\begin{equation}
x^{2}g_{LT}^{(U)}(x,Q^{2}) \approx {r_{5}(0)}\cdot 10^{-4} 
\left({0.001 \over x }\right)^{0.4}
\label{eq:18}
\end{equation}
It corresponds to spin symmetry 
\begin{equation}
A_{2} \approx 6\cdot 10^{-4}\cdot r_{5}(0){m_{p} \over \sqrt{Q^{2}}} 
\label{eq:19}
\end{equation}
which is approximately flat in the region of $x =(10^{-3}--10^{-4})$, 
cf. the $x$-dependence of tensor asymmetry in \cite{NNSchaefer}.

The steep rise (\ref{eq:18}) of $g_{LT}^{(U)}$ can not go forever as it
would conflict the unitarity bounds. The same is true of the experimentally
observed rise of unpolarized structure functions. The scale of unitarity 
effects is set by the ratio of diffractive to nondiffractive DIS 
\cite{NZ94,BaroneUnit}. One could evaluate higher order unitarity effects
consistently by the technique developed in \cite{ZakharovLPM}, here we 
only notice that higher order unitarity  corrections are 
known to play marginal role in related nuclear shadowing in DIS on even
heaviest nuclei \cite{Barone}.

Above we focused on the two-pomeron cut contribution
which dominates at very small $x$. The related contribution from 
secondary reggeon-pomeron cut will be of the form 
$$
x^{2}g_{LT}^{(R\Pom)} \propto 
\left({1 \over x}\right)^{\delta_{g}+\alpha_{R}-1}\, .
$$
Because of a large spin-flip coupling of secondary reggeons (for the
review see \cite{Buttimore}), this subleading term can well dominate 
the unitarity correction at moderately small $x$. 
It will definitely dominate the small-$x$ behaviour of proton-neutron 
difference $g_{LT}^{p}-g_{LT}^{n}$. It could eventually
be evaluated with the further progress in QCD modeling of reggeon
effects in diffractive DIS \cite{SchaeferDIS98}. With the reference to
reggeon studies in \cite{SchaeferDIS98}, here we only emphasize that
$g_{LT}^{(R\Pom)}(x,Q^{2})$ is a scaling function of $Q^{2}$.

As a reference value for the comparison with our small-$x$ result, we 
show in fig.~3 the so-called Wandzura-Wilczek (WW) result \cite{WW}
\begin{equation}
x^{2}g_{LT}^{WW}(x,Q^{2}) = x^{2}\int_{x}^{1} {dy \over y}
g_{1}(y,Q^{2})\, .
\label{eq:20}
\end{equation}
The standard parameterizations and QCD ladder estimate \cite{ErmolaevG1}
give $g_{1}(x,Q^{2}) \sim \left({1\over x}\right)^{\delta_{1}}$ with the
exponent $\delta_{1} \sim$ 0-0.5. Then $x^{2}g_{LT}^{WW}(x,Q^{2})$ would
vanish at small $x$ as $x^{2-\delta}$. To the extent that they are fitted 
to the same experimental data, all available parameterizations of $g_{1}$
give approximately the same WW integral, our WW curve shown in fig.~3 is
for the parameterization \cite{Stirling}. Our diffractive mechanism takes
over at $x \lsim 10^{-3}$. 

The uncertain status of the WW relation has been much discussed in the 
literature \cite{Jaffe,Anselmino,Reya}. The WW relation has 
never been supposed to, and evidently does not, hold in presence of such a 
diffractive component of $g_{LT}$. However, in the spirit of duality sum 
rules one may still hope that full $g_{LT}$ minus our diffractive 
component minus the $R\Pom$ cut contribution has the required 
superconvergence properties and one can hypothesize that the 
WW relation is applicable to $g_{LT}-g_{LT}^{(U)}$. In other words, it 
is tempting to identify our unitarity effect $g_{LT}^{(U)}$ with the 
long sought deviation from the WW relation and to hypothesize that WW 
relation would hold approximately at large and moderate $x$ where the 
diffractive component is numerically small. This seems to be the case 
in the experimentally studied range $x \gsim 0.01$ \cite{SLAC}.  

As well known, $g_{LT}(x,Q^{2})$ does not admit any obvious parton
model interpretation.
The OPE content of unitarity corrections to structure functions deserves
a dedicated study. We only notice that in close similarity to leading 
twist unpolarized diffractive structure function \cite{NZ94}, the upper
blob in diagrams of fig.~2 receives a substantial contribution from 
large and moderate transverse distances. So to say, in the quark loop
we are way along the light cone, but finite, not $1/Q$, transverse distance
from the light cone. 

\section{Unitarity driven $g_{LT}^{(U)}$ from vector meson production?}

We wish to point out that diffractive vector mesons offer a direct 
experimental window at effective $r_{5}(0)$ which sums up the elastic
and inelastic rescattering contributions. Evidently, unitarity diagrams
of figs.~1,2 do contribute to $\gamma^{*}_{\nu} p_{\lambda} \to 
V_{\mu}p'_{\rho}$ too. Our diffractive mechanism would give rise
to finite $ A^V_{0-{1\over2},+1+{1\over 2}}({\bf\Delta}=0)$,
whereas in two-parton ladder ($=$ one pomeron exchange)
$ A^V_{0-{1\over2},+1+{1\over 2}}({\bf\Delta})\propto {\bf\Delta}^{2}$.
The marginal difference from the above evaluation of diffractive $g_{LT}$
is  emergence of  $q\bar{q}V$ vertex instead of one of the pointlike QED 
$q\bar{q}\gamma$ vertices and that the two gluon distributions in vector
meson production are skewed differently than in the calculation of 
$g_{LT}^{(U)}$,
but the ratio of diffraction driven SCHNC LT and SCHC conserving 
dominant one-pomeron amplitudes can be worked out. What counts is that
the lower blob of the diagram is identical to that in the case 
of $g_{LT}^{(U)}$. 
For the experimental determination of  $A_{0-{1\over2},+1+{1\over 2}}
({\bf\Delta}=0)$ one needs to isolate the polarization dependence of 
production of longitudinally polarized vector mesons by circular polarized 
photons on transverse polarized targets, which is doable experimentally
because decays of vector mesons are self analyzing. Notice, that unitarity 
considerations in section 3 were quite general and did not require an 
applicability of pQCD, finite $Q^{2}$ was only needed to have longitudinal
photons. In the case of vector mesons, nonvanishing $A^V_{0-{1\over2},
+1+{1\over 2}}({\bf\Delta}=0)$ is possible not only with virtual
photons in polarized DIS but 
also for circular polarized 
real photons. Although real photoproduction of $\rho$ and $\phi$ mesons
will be utterly nonperturbative process, the cross sections are large, 
high luminosity external real photon beams can readily be produced either 
by laser backscattering or coherent bremsstrahlung in crystals, and 
vector meson production seems to be an ideal testing ground for existence
of diffraction driven $g_{LT}^{(U)}$. Because much of the dependence on the 
model of the vector meson wave function would cancel in the ratio
of SCHNC spin-flip and SCHC non-flip amplitudes, even nonperturbative
real photoproduction of vector mesons would provide useful constraints 
on $r_{5}(0)$.  

\section{Can the Burkhardt-Cottingham sum rule be salvaged?}

Our diffraction driven contribution to $g_{LT}(x,Q^{2})$ rises at small 
$x$ faster than $g_{1}(x,Q^{2})$. Incidentally, the diffractive mechanism
does not contribute to the difference helicity of amplitudes which gives
rise to the spin asymmetry $A_{1}$. Consequently, at small $x$ the 
diffraction driven $g_{LT}^{(U)}$ is dominated by $g_{2}$. The resulting 
small-$x$ rise of $g_{2}$ invalidates the superconvergence assumption 
behind the derivation of the BC sum rule \cite{BC}. There were 
suggestions reviewed in \cite{Jaffe,Reya} that the BC sum rule might 
be salvaged because the residues of pomeron cuts might vanish at large 
$Q^{2}$. This is not the case with our diffractive  
$g_{LT}^{(U)}(x,Q^{2})$ which is a manifestly scaling function of $Q^{2}$. 
To this end we recall that unpolarized diffractive DIS is a well 
established scaling phenomenon (\cite{GNZcharm,NZ94,GNZ95}, for the 
corresponding phenomenology and review of the HERA data on diffractive 
DIS see \cite{BGNPZ98,Ellis}).

\section{Impact of diffraction driven $g_{LT}^{(U)}$ to extraction
of $g_{1}$ from longitudinal asymmetry}

As well known there is a $\propto {1\over Q^{2}}$ correction from
$g_{2}$ in the extraction of the small-$x$ spin structure function 
$g_{1}$ from asymmetry $A_{1}$:
\begin{eqnarray}
A_{1}={1 \over F_{1}(x,Q^{2})}\left\{g_{1}(x,Q^{2})-{4m_{p}^{2}x^{2} 
\over Q^{2}}g_{2}(x,Q^{2})\right\}  \nonumber \\
= 
{1 \over F_{1}(x,Q^{2})}\left\{g_{1}(x,Q^{2}(1+{4m_{p}^{2}x^{2} 
\over Q^{2}}) - {4m_{p}^{2} 
\over Q^{2}}x^{2}g_{LT}(x,Q^{2})\right\}
\label{eq:21}
\end{eqnarray}
The small-$x$ growth of $x^{2}g_{LT}^{(U)}(x,Q^{2})$ invalidates the 
common lore assumption that this correction can be neglected at low $x$.
Quite to the contrary, it is not dissimilar to, or even somewhat faster
than, the usually discussed
small-$x$ rise of $g_{1}(x,Q^{2})$ and this term can not be neglected 
off hand. Our conservative estimate (\ref{eq:18}) suggests that it is
small, though.

\section{ Summary and conclusions}

We have shown how  $s$-channel helicity nonconserving LT interference 
in diffractive DIS in conjunction with the pomeron spin-flip in 
diffractive nucleon-nucleon scattering gives rise to a steep rise
eq.~(\ref{eq:1}) of the spin structure function $g_{LT}(x,Q^{2})$ at
small $x$. The transverse spin asymmetry considered in this paper and 
tensor spin asymmetry discussed earlier in \cite{NNSchaefer} fall into 
a broad family of unitarity (diffraction) driven spin effects which, 
in the opposite to the common wisdom, persist in high energy and/or 
small-$x$ limit (the work on straightforward extension of the above 
to related spin structure functions in DIS off photons is in progress). 
The rate of rise of diffraction driven $g_{LT}^{(U)}(x,Q^{2})$ is related 
to that of other experimental observable - the unpolarized gluon 
structure function of the target proton or, still better, to the 
experimentally measurable $x_{\Pom}$ dependence of unpolarized diffractive 
structure function. Whether the diffraction driven $x^{2}g_{LT}^{(U)}
(x,Q^{2})$ is numerically large or small is not an issue, the crucial
point is that the found rate of the small-$x$ rise of $g_{LT}(x,Q^{2})$ 
invalidates the superconvergence assumptions behind the 
Burkhardt-Cottingham sum rule and behind the Wandzura-Wilczek relation. 
There exists an interesting possibility of testing the existence of 
diffractive mechanism for
$g_{LT}$ in vector meson production by circular polarized real photons 
on transverse polarized proton targets, which deserves dedicated study. 
 \\

{\bf Acknowledgments:} Our warm thanks go to B.G.Zakharov for inspiration
and much insight at early stages of this work. We are grateful to
B.Ermolaev and I.Ginzburg for useful discussions. IPI thanks Prof. J.Speth 
for the hospitality at the Institut f. Kernphysik of Forschungszentrum 
J\"ulich. The work of IPI has been partly supported by RFBR, the work 
of A.V.P. was supported partly by the US DOE grant DE-FG02-96ER40994.
\pagebreak\\

{\bf Figure captions.}\\

\noindent
{\bf Fig.~1.} The unitarity diagram with (a) `elastic' $pX$ intermediate
state and (b) `inelastic' $p^{*}X$ intermediate state with excitation
of the proton to resonances or low-mass continuum $p^{*}$.\\

\noindent
{\bf Fig.~2.} (a) The QCD model for unitarity diagram of Fig.~1. The
lower blobs are related to unintegrated gluon structure function of the 
proton and contain the pomeron spin-flip amplitude.
(b) The pQCD Feynamn diagram content of the shaded upper blobs. 
\\

\noindent
{\bf Fig.~3}. The conservative estimate assuming $r_{5}(0)=-0.1$ for
the unitarity driven $x^{2}g_{LT}^{(U)}(x,Q^{2})$ vs. the
expectation from WW relation (the dashed curve) for GS \cite{Stirling} 
parameterization for $g_{1}(x,Q^{2})$.  The difference between 
the curves for $Q^{2}=5$ GeV$^{2}$ (diamonds) and $Q^{2}=100$ GeV$^{2}$
(triangles) is due to the charm contribution at large $Q^{2}$.
    

\begin{thebibliography}{299}

\bibitem{BC} 
H.Burkhardt and W.N.Cottingham, {\sl Ann. of Phys. (USA)},
{\bf 56}, 453 (1970) 

\bibitem{Zakharov}
B.G. Zakharov, {\sl Yad. Fiz.} {\bf 49}, 1386 (1989); {\sl  Sov. J.
Nucl.
Phys.} {\bf 49}, 860 (1989); B.Z. Kopeliovich and B.G. Zakharov,
{\sl Phys. Lett.} {\bf B226}, 156 (1989).

\bibitem{NPLT} 
N.N. Nikolaev, A.V. Pronyaev and B.G.Zakharov, e-Print Archive: 
hep-ph/9812212, {\sl Phys. Rev.} {\bf D} (1999) in print.

\bibitem{KNZ98} 
E.V.Kuraev, N.N.Nikolaev and B.G.Zakharov, 
{\sl JETP Lett.} {\bf 68},  667 (1998)

\bibitem{IN99} 
I.P.Ivanov and N.N.Nikolaev,  {\sl JETP Letters} {\bf 69}, 268 (1999)

\bibitem{Gribov} 
V.N. Gribov, {\sl Sov. J. Nucl. Phys.} {\bf 5}, 138 (1967); {\sl Yad.
Fiz.}
{\bf 5}, 197 (1967).

\bibitem{NNSchaefer} 
N.N. Nikolaev and W. Schafer, {\sl Phys. Lett.} {\bf B398}, 245 (1997) 
Erratum-ibid. {\bf B407}, 453 (1997).

\bibitem{CloseKumano} 
F.E. Close and S. Kumano, {\sl Phys. Rev.} {\bf D42}, 2377 (1990). 

\bibitem{Karnakov}  
V.M.Karnakov, {\sl Yad. Fiz.} {\bf 37}, 1258 (1983); {\sl Sov. J.
Nucl.
Phys,} {\bf 37}, 748 (1983)

\bibitem{AGK} 
V.A. Abramovskii, V.N. Gribov, O.V. Kancheli, {\sl Yad. Fiz.} {\bf 18},
595 (1973); {\sl Sov. J. Nucl. Phys.} {\bf 18}, 308 (1974).

\bibitem{Ioffe} 
B.L.Ioffe, V.A.Khoze and L.N.Lipatov, Hard Processes, North Holland 
(Amsterdam-Oxford-New York-Tokyo), 1984.

\bibitem{BFKL} 
E.A.Kuraev, L.N.Lipatov and S.V.Fadin,
{\sl Sov. Phys. JETP} {\bf 44} (1976) 443;
{\sl Sov. Phys. JETP} {\bf 45} (1977) 199;
L.N.Lipatov, {\sl Sov. Phys. JETP} {\bf 63} (1986) 904;
L.N. Lipatov, {\sl Phys. Rept.} {\bf 286}, 131 (1997);
V.S. Fadin and L.N. Lipatov, {\sl Phys. Lett.} {\bf B429}, 127 (1998);

\bibitem{ErmolaevVal} 
B.I. Ermolaev, S.I. Manaenkov and M.G. Ryskin, {\sl Z. Phys.} {\bf
C69},
259 (1996); R. Kirschner and L.N. Lipatov, {\sl Nucl. Phys.} {\bf
B213},
122 (1983).

\bibitem{ErmolaevG1} 
J. Bartels, B.I. Ermolaev and  M.G. Ryskin, {\sl Z. Phys.} {\bf C70},
273 (1996); {\sl Z. Phys.} {\bf C72}, 627 (1996).

\bibitem{ErmolaevG2} 
B.I. Ermolaev and S.I. Troian, e-Print Archive: hep-ph/9703384.

\bibitem{ErmolaevF3} 
B. Ermolaev, R. Kirschner and L. Szymanowski, 
e-Print Archive: hep-ph/9806439.

\bibitem{Jaffe} 
R.L.Jaffe, {\sl Comm. Nucl. Part. Phys.} {\bf 19}, 239 (1990)

\bibitem{Anselmino} 
M. Anselmino, A. Efremov and E. Leader, {\sl Phys. Rept.} {\bf 261},
1 (1995) Erratum-ibid. {\bf 281}, 399-400 (1997)

\bibitem{Reya} 
B. Lampe and E. Reya, {\bf MPI-PHT-98-23}, 
e-Print Archive: hep-ph/9810270.

\bibitem{Heimann} 
R.L. Heimann, {\sl Nucl. Phys.} {\bf B64}, 429 (1973).

\bibitem{GribovShad} 
V.N.Gribov, {\sl Sov.Phys. JETP} {\bf29} (483) 1969

\bibitem{GNZcharm} 
M.Genovese, N.Nikolaev  and B.Zakharov,
{\sl Phys. Lett.} {\bf B378}, 347 (1996);{\bf B380},  213 (1996)

\bibitem{BGNPZ98} 
M.Bertini, M. Genovese, N.N. Nikolaev, A.V. Pronyaev and B.G.
Zakharov, 
{\sl Phys. Lett.} {\bf B422}, 238 (1998).

\bibitem{NPZslope}
N.N. Nikolaev, A.V. Pronyaev and B.G. Zakharov, {\sl JETP Lett.} {\bf 68},
634 (1998)
 
\bibitem{NZ94} 
N.N.Nikolaev and B.G.~Zakharov, {\it Z. Phys.}
{\bf C53},  331 (1992); {\sl JETP}
{\bf 78}, 598 (1994);
{\sl Z. Phys.} {\bf C64} (1994) 631.


\bibitem{GNZ95} 
M.Genovese, N.Nikolaev  and B.Zakharov,
 {\sl JETP} {\bf 81} 625 (1995);
 {\sl JETP} {\bf 81},  633 (1995).

\bibitem{Radyushkin}
A.V. Radyushkin, {\sl Phys. Lett.} {\bf B385}, 333 (1996)

\bibitem{Ji}
X. Ji, {\sl Phys. Rev.} {\bf D55}, 7114 (1997)

\bibitem{Buttimore} 
N.H. Buttimore et al., e-Print Archive: hep-ph/9901339.
 
\bibitem{NNN}
N.N. Nikolaev, {\sl Surveys High Energ. Phys.} {\bf 7}, 1 (1994);
{\sl Int. J. Mod. Phys.} {\bf E3}, 1 (1994), Erratum-ibid. {\bf E3},
995 (1994).

\bibitem{BaroneUnit} 
V. Barone, M. Genovese, N.N. Nikolaev, E. Predazzi and B.G. Zakharov,
{\sl Phys. Lett.} {\bf B326}, 161 (1994)

\bibitem{ZakharovLPM}
B.G. Zakharov, {\sl Phys. Atom. Nucl.} {\bf 61}, 838 (1998); 
{\sl Yad. Fiz.} {\bf 61}, 924 (1998) 

\bibitem{Barone}
V. Barone et al., {\sl Z. Phys.} {\bf C58}, 541 (1993)

\bibitem{SchaeferDIS98}
W.Sch\"afer, Deep Inelastic Scattering and QCD - DIS'98. Proceedings 
of 6th International Workshop, Brussels, Belgium, 4-8 April 1998,
editors Gh, Corenmans and R.Roosen, World Scientific, pp. 404-407.

\bibitem{WW}
S.Wandzura and F.Wilczek, {\sl Phys. Lett.} {\bf B72}, 195 (1977)

\bibitem{Stirling}
T. Gehrmann and  W.J. Stirling, {\sl Phys. Rev.} {\bf D53}, 6100 (1996)


\bibitem{SLAC}
P.L.Anthony et al., {\bf SLAC-PUB-7983} (1999). 
E-print Archive: hep-ex/9901006

 
\bibitem{Ellis}
J. Bartels, J. Ellis, H. Kowalski and M. Wusthoff, {\bf CERN-TH-98-67}, 
hep-ph/9803497; J. Bartels and C. Royon, hep-ph/9809344. 
\pagebreak\\
\end{thebibliography}
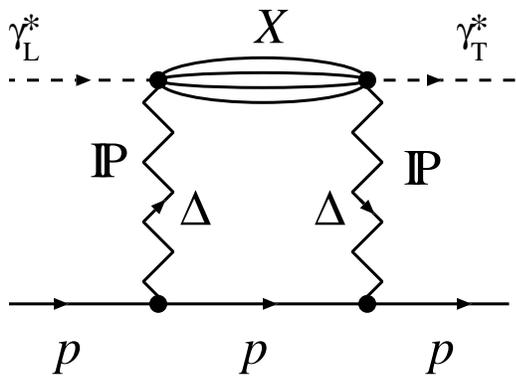
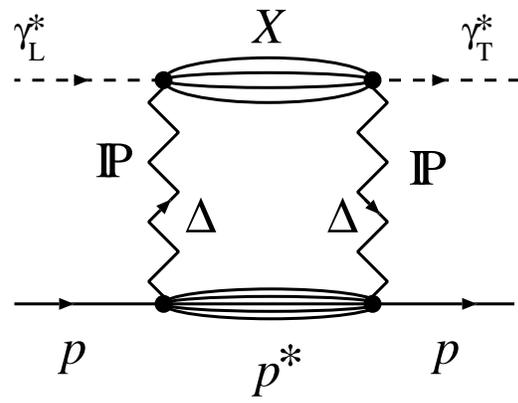
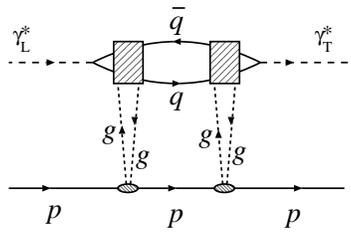
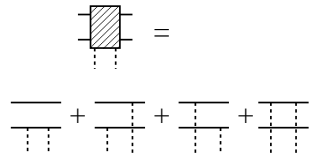
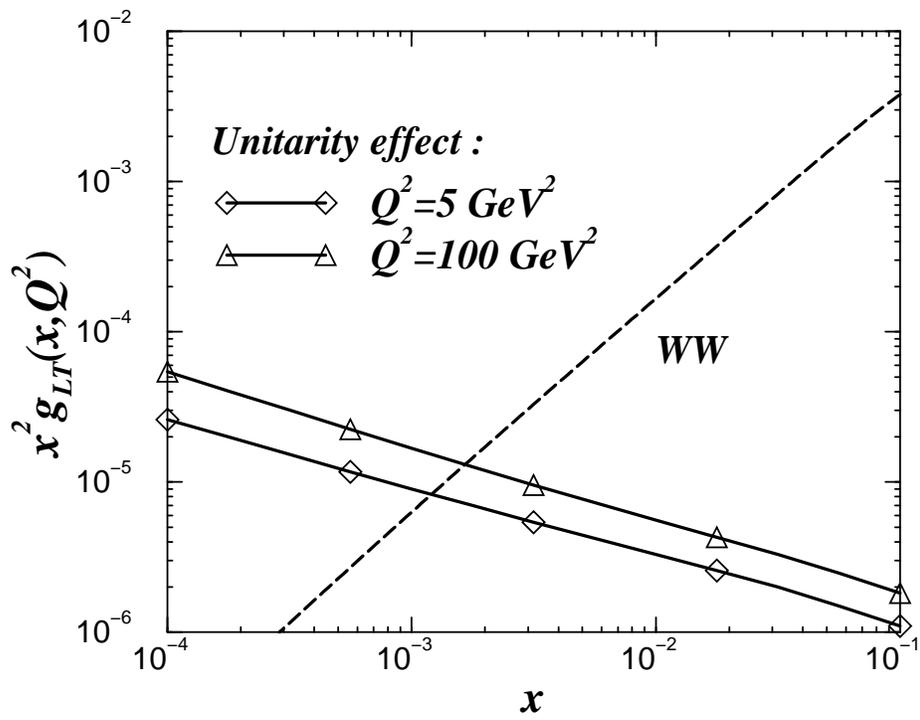
\end{document}